\documentclass{article}
\usepackage[a4paper, total={6in, 8in}]{geometry}
\usepackage{calrsfs}
\usepackage{hyperref}
\usepackage{color}
\usepackage{float}
\usepackage{esvect}
\usepackage{booktabs}
\usepackage{siunitx}
\usepackage{authblk}
\usepackage{calrsfs}
\usepackage{amsmath}
\usepackage{amssymb}
\usepackage{braket}
\usepackage[version=4]{mhchem}
\usepackage{mathtools}          

\usepackage{graphicx}
\usepackage{dcolumn}
\usepackage{graphicx}
\usepackage{bm}
\usepackage{sectsty}

\sectionfont{\fontsize{12}{15}\selectfont}

\begin{document}

\title{Experimental verification of the temperature coefficient of resistivity}
\author[1]{\small Robert D. Polak}
\author[1]{\small Michael R. Harris}
\author[1]{\small Kiet A. Nguyen}
\author[1]{\small Anthony Kearns}
\affil[1]{\small Department of Physics, Loyola University Chicago, Chicago, IL 60660, USA}




%


%
%
%


\date{}

\maketitle

\section*{}
We have created an experimental procedure for determining the temperature coefficient of resistivity, $\alpha_R$, for introductory physics laboratories. As in the procedure from Henry [1], this method examines the relationship between temperature and resistivity to establish $\alpha_R$ within 10\% of the accepted value.
\\

Electrical resistivity, $\rho$, varies with temperature according to:
\begin{equation}
    \rho = \rho_o (1+ \alpha_R (T - T_o))
\end{equation}
where $\rho_o$ is the resistivity for a given temperature $T_o$, $T$ is the temperature of the material, and $\alpha_R$ is the temperature coefficient of resistivity. For a wire of length, $L$, and cross-sectional area, $A$, the resistance of a wire, $R$, is defined accordingly as
\begin{equation}
R= \rho \frac{L}{A} \ .    
\end{equation}

While resistance will increase as a product of both increased length and resistivity, the increase in length provides a negligible increase in resistance. This is evident from observing that the thermal coefficient of resistivity is approximately two orders of magnitude larger than the coefficient of thermal expansion. As such, the change in resistivity is primarily responsible for the increase in resistance. Hence, $R$ will vary similarly with
\begin{equation}
    R=R_o(1+\alpha_R(T-T_o))
\end{equation}
where $R_o$ is the resistance of the wire at a temperature $T_o$.


By applying a current through the wire, its temperature will also vary as a result of Joule heating and the resistance of the wire can be measured based on a given current, I, and difference in voltage, $\Delta V$, by
\begin{equation}
    R=\frac{\Delta V}{I} \ .
\end{equation} 
Hence, by measuring the resistance as a function of temperature, we can determine $\alpha_R$ by plotting $R$ vs. $T-T_o$ and performing a linear fit using Eq. (3).

To perform the experiment, we created a closed circuit (see Fig. 1) in which a carbon steel wire [2] is suspended under tension above a surface, as in most stringed instruments. Two digital multimeters were used to record the voltage across and current through a 0.016 inch (0.406 mm) diameter, 40-cm long wire.
Temperature measurements were taken using liquid crystal thermometers [3] placed in thermal contact with the wire by fastening them to the wire with an adhesive backing. Two thermometers were used, one ranging from $14-31 ^o C$ and the other from $32-49 ^o C$, to provide an overall temperature range of $14-49 ^o C$. For the most accurate temperature readings, we found it is essential to avoid all contact with the thermometers during the experiment. To collect the data, we applied different currents to the wire, ranging from $0.2 A$ to $1.0 A$. We used a BK Precision 1787B power supply that allows for digital control of the current. We found that having initial steps of $0.2 A$ and later reduced to $0.1 A$ created consistent temperature changes in the wire of $2-3 ^o C$. The experiment proved much more difficult to complete using an analog power supply because of the difficulty in creating the precise changes in current needed to have a well formed data set. After allowing around 30 seconds for the system to reach thermal equilibrium, the recorded temperature of each trial was given as the uppermost visible temperature reading of the liquid crystal thermometer, as seen in Fig. 2. We recorded the temperature of, current through and voltage across the wire, and calculated its resistance using Eq. (4).

By graphing the resistance of the wire as a function of $T-T_o$, where  $T_o$ is the temperature of the wire with the lowest current applied, we can then apply a linear fit to the data with the y-intercept yielding $R_o$ and slope giving $R_o \alpha_R$, according to Eq. (3). Figure 4 shows example experimental results with the fit giving $R_o=0.744 \Omega$ and $\alpha_R = 0.0039 K^{-1}$. This is within 5\% of the accepted value of $\alpha_R = 0.0041 K^{-1}$ [4]. Repeated experiments found these results to be reproducible with $\alpha_R$ consistently measured within 10\% of the accepted value.

To get the best results, we found that the resistance of the wire should be at least $0.3\Omega$ to allow for accurate resistance measurements. Furthermore, the wires need to be thick enough to support the thermometers. As such, we achieved the best results when using steel as opposed to other materials with lower resistivity and tensile strength, such as copper and aluminum.

This experiment can be completed in less than 2 hours and uses equipment that is commonly present in a typical introductory physics lab, with the exception of relatively low cost supplies such as music wire and liquid crystal thermometers. It also reinforces key ideas from introductory physics such as conservation of energy, where electrical energy becomes thermal energy, Ohm's Law, and the temperature dependence of resistance.

\section*{References}
1. D. Henry, ``Resistance of a wire as a function of temperature", The Physics Teacher 33, 96-97 (1995) \url{https://doi.org/10.1119/1.2344149}\\
2. Precision Brand Music Wire (0.016 inch diameter); UPC. No. 21016 \\
3. TelaTemp reversible LCT strip model 416-2 ($14-31 ^o C$) and 416-3 ($32-49 ^o C$) \\
4. S. Yafei , N. Dongjie, and S. Jing,  4th IEEE Conference on Industrial Electronics and Applications, 368-372 (2009).

\begin{figure}[h]
\centering
\includegraphics[scale=.4]{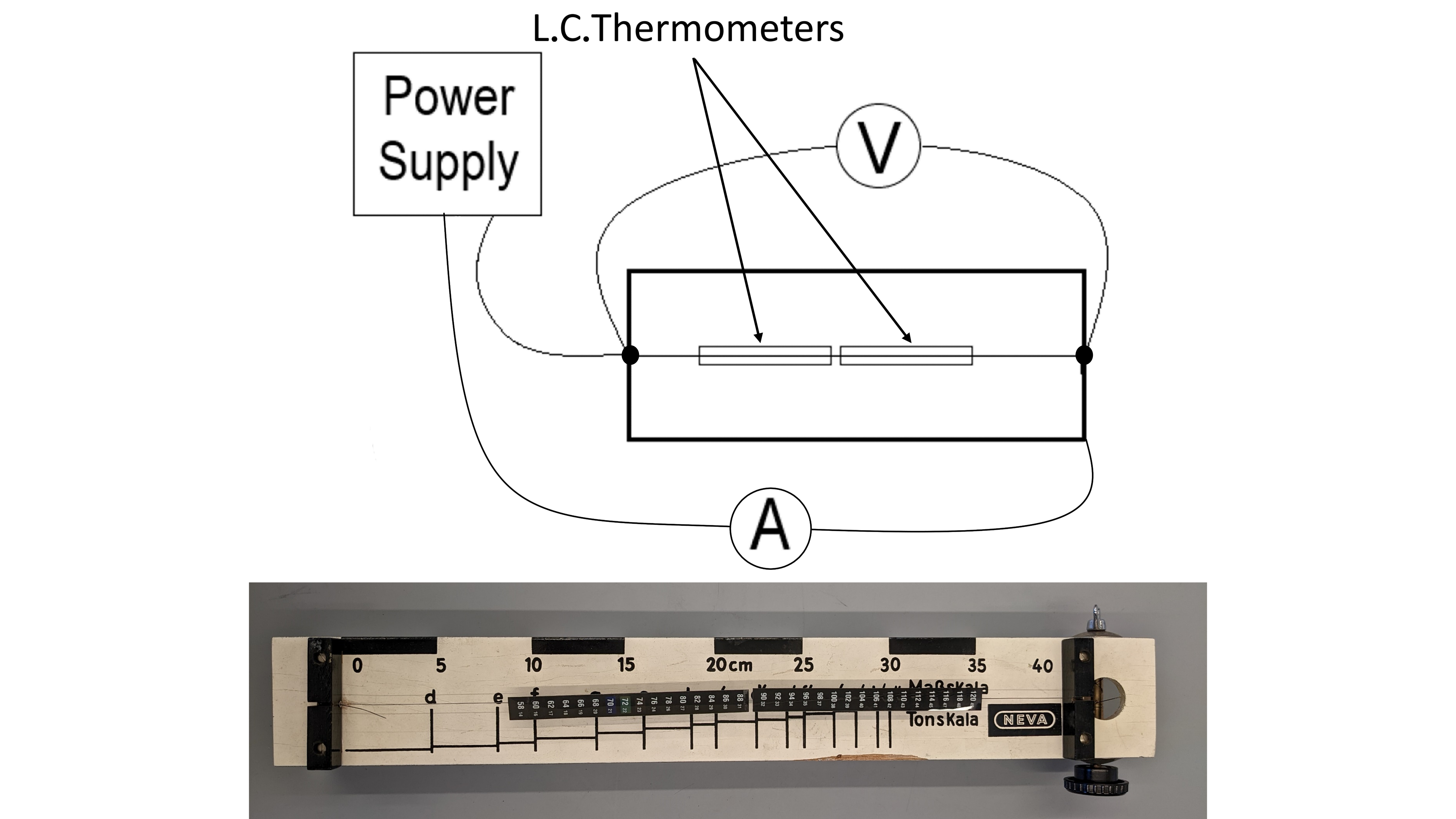}
\caption{The experimental setup (bottom) with the wire suspended above the box, and a circuit diagram (top). The circuit consists of two multimeters set up to measure current and voltage and a digital power supply. The liquid crystal thermometers are affixed to the top of the wire.}
\end{figure}
\begin{figure}[h]
\centering
\includegraphics[width=6cm, height=9cm,angle= 0]{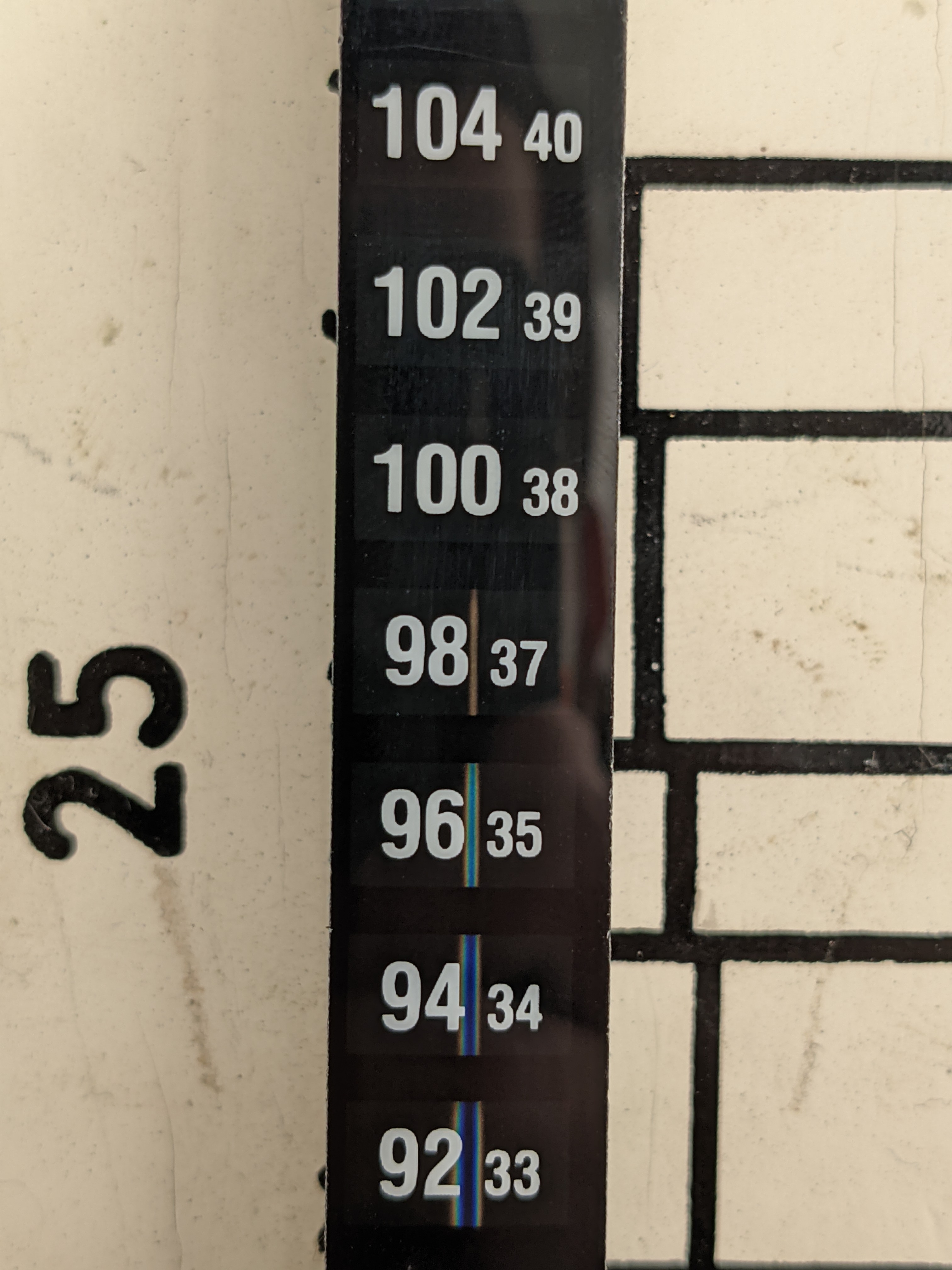}
\caption{An example of a temperature reading corresponding to $37^{\circ} C$ ($98^{\circ} F$), according to the procedure of reading the uppermost visible indication of the liquid crystal thermometer (the yellow bar spanning $37^{\circ} C$).}
\end{figure}
\begin{figure}[h]
\centering
\includegraphics[width=8cm, height=6cm, angle= 0]{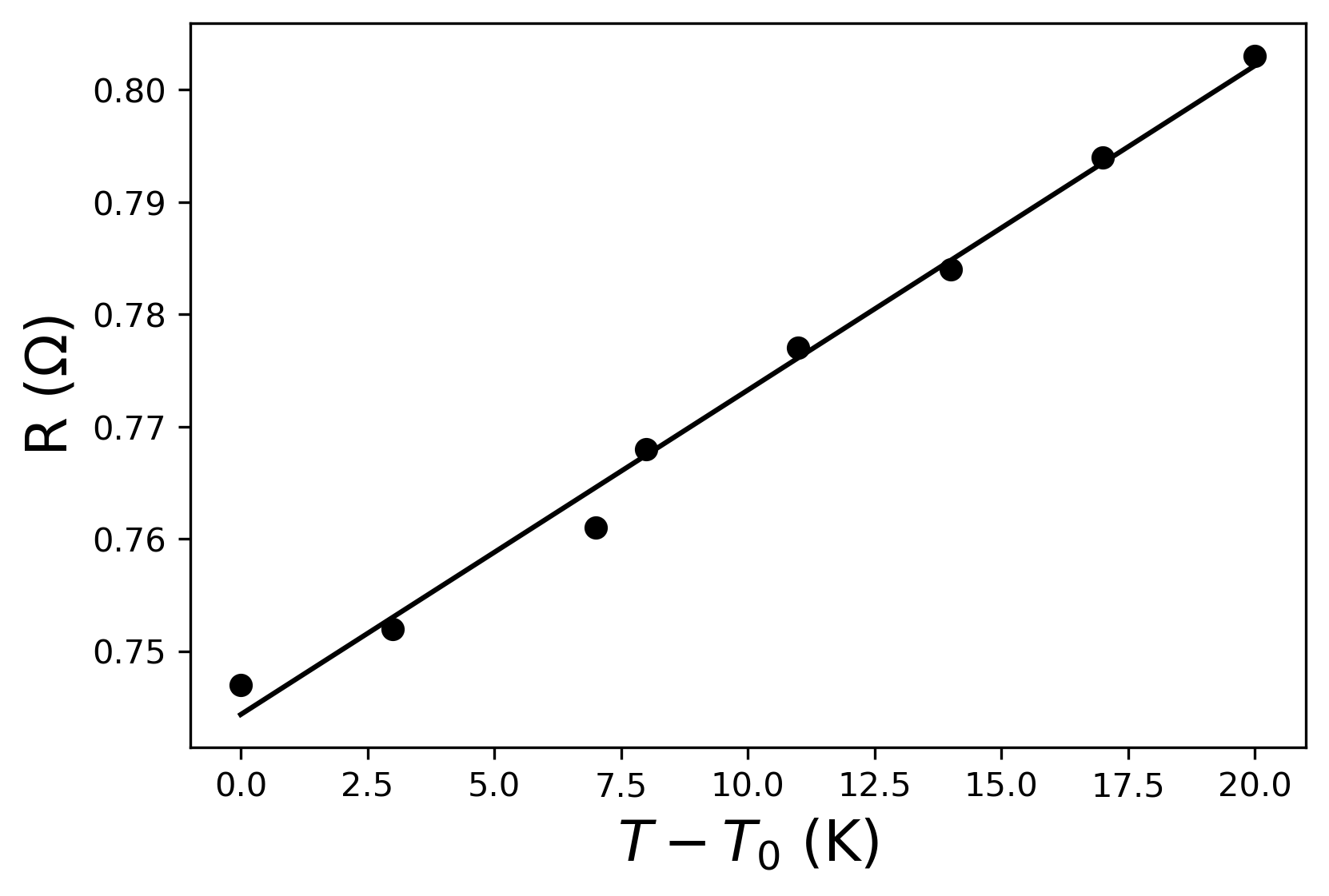}
\caption{Plot of resistance against the associated change in temperature for currents from 0.2 $A$ to 1.0 $A$. A linear fit of the data yields $\alpha_R = 0.0039 K^{- 1}$.}
\end{figure}


\end{document}